%% file: arxiv.tex
\documentclass{article}
\usepackage{arxiv}
\usepackage{booktabs,tabularx}
\usepackage{xcolor}
\usepackage{colortbl}
\usepackage{array}
\usepackage{multirow}
\usepackage{enumitem}
\usepackage{graphicx}
\usepackage{subcaption}
\usepackage{arydshln}
\usepackage{amssymb}
\usepackage{enumitem}
\usepackage{tcolorbox}
\tcbuselibrary{breakable}
\usepackage[multiple]{footmisc}
\usepackage{amsmath}
\usepackage{hyperref}
\usepackage[numbers,sort&compress]{natbib}
\usepackage{fancyhdr}       

\input{macros}
\AtBeginDocument{
  \fancypagestyle{firstpagestyle}{
    \fancyhead{}%
    \fancyfoot[C]{}%
  }
  \fancyhf{}
\pagestyle{fancy}
\fancyhf{}

\newcommand{\runningtitle}{Is This AI? Longitudinal Analysis of Strategies Used for AI Detection on Two Subreddits}
\newcommand{\runningauthors}{{Yeung et al.}}
\setlength{\headwidth}{\textwidth}

\fancyhead[L]{\sffamily\footnotesize\ifodd\value{page}\runningtitle\fi}
\fancyhead[R]{\sffamily\footnotesize\ifodd\value{page}\else\runningauthors\fi}
\fancyfoot[C]{}

  \fancyfoot[C]{}%
}
\makeatother

\title{Is This AI? Longitudinal Analysis of Strategies Used for AI Detection on Two Subreddits}

\usepackage{authblk}

\author[1]{%
    {Christina Yeung}%
}
\author[1]{%
    {Galen Weld}%
}
\author[2]{%
    {Jaron Mink}%
}
\author[1]{%
    {Franziska Roesner}%
}
\affil[1]{University of Washington}
\affil[2]{Arizona State University}
\affil[ ]{\small \texttt{\{cyeung3, gweld\}@cs.washington.edu, jaron.mink@asu.edu, franzi@cs.washington.edu}}

\begin{document}

\maketitle

\begin{abstract}
\input{sections/00_abstract}
\end{abstract}

\input{sections/10_introduction}
\input{sections/20_related}

\input{sections/30_methods}
\input{sections/40_results}

\input{sections/500_discussion}
\input{sections/600_futurework_limitations}

\input{sections/700_conclusion}
\section*{Acknowledgments}

We are very grateful to the following people for their feedback and conversations in the development of this work and paper: Natalie Grace Brigham, Maddie Burbage, Ian Chang, Elise Dorough, Joe Eckert, Yael Eiger, Kshitish Ghate, Blaine Hoak, Rachel Hong, David Kohlbrenner, Rachel McAmis, Chris Nagle, Basia Radka, Les Sessoms, Michael Tompkins, Henry Wong. Thank you especially to Natalie Grace Brigham for help with qualitative coding.
This work was supported in part by the U.S. National Science Foundation under Award \#2114230 and by a Microsoft Grant for Customer Experience Innovation.

\bibliographystyle{ACM-Reference-Format}
\bibliography{refs}

\appendix
\input{sections/800_statements}
\input{sections/X_appendix}

\end{document}

%% file: macros.tex
\newcommand{\franzi}[1]{\textsf{\textbf{\small\textcolor{orange}{[FR: #1]}}}}
\newcommand{\jaron}[1]{\textsf{\textbf{\small\textcolor{purple}{[JM: #1]}}}}
\newcommand{\tina}[1]{\textsf{\textbf{\small\textcolor{blue}{[TY: #1]}}}}

\newcommand{\header}[1]{{\smallskip\noindent\textbf{#1.}}}

\newcommand{\pquote}[1]{\textit{``#1''}}

%% file: sections/00_abstract.tex
As AI-generated content (e.g., “slop”) becomes more prevalent online, people are developing strategies to attempt to identify it (or, conversely, to gain confidence that something is not AI-generated). What strategies are people using, and how are they changing over time as generative AI models themselves change? In this work, we catalog and analyze 2 years and 8 months of the AI detection strategies discussed by users of two popular Reddit communities (r/isthisAI and r/RealOrAI) that use the wisdom of crowds to identify AI-generated media. Through a mixed-method analysis of 13,098 posts and 222,060 comments within these communities, we catalog and analyze the prevalence of 12 AI-detection strategies, including examining fine-grained physical details, recognizing trends in AI-created content, and the assumptions people make about what models are capable of producing. Furthermore, we find that these strategies and mental models shift over time in accordance with changing AI capabilities and in response to online social trends. 
By systematically cataloging users' AI detection strategies, we lay the groundwork for user-facing guidance and future research.

%% file: sections/10_introduction.tex
\section{Introduction}
\label{sec:intro}

The last few years have seen tremendous growth in generative AI in terms of model capabilities, the accessibility of these tools to end-users, and the proliferation of AI-generated content online, including text, images, audio, and videos. Though these capabilities and tools are exciting for some use cases and some users, they also raise myriad concerns, including 
the potential for ``deep fakes’’ to spread political or other disinformation~\cite{naitali2023deepfake}, interpersonal abuse (e.g., non-consensual generated intimate imagery)~\cite{gibson2025analyzing}, and the proliferation of low-quality content on the web and social media (often called ``AI slop’’)~\cite{hoffman2024slop, shaib2025measuring}.

As a partial mitigation to some of these concerns, it may be helpful for users to be able to distinguish genuine content (e.g., human-generated text or unaltered photos/videos) from AI-generated content online. For example, correctly identifying AI-generated content can help users avoid political misinformation or scams; on the other hand, believing something AI-generated is real may contribute to the spread of misinformation. However, as generative AI techniques become increasingly sophisticated, it has also become increasingly difficult for people to tell the difference just by looking: for instance, an  AI-generated video clip of bunnies jumping on a trampoline went viral in 2025, fooling people into thinking that it was authentic~\cite{thomas_ai_2025}.

There are several approaches to help remedy the difficulty of identifying AI-generated content, though each with their limitations. 
AI watermarking schemes are promising, yet are currently easily bypassed by simply using a tool that does not watermark its output \cite{christ2024undetectable, zhao2025sokwatermarkingaigeneratedcontent,text_watermarking, dathathri2024scalable,hussain2021adversarial}.
Furthermore, automated AI detection tools are subject to an ever-escalating `arms race' between people generating AI content and those seeking to detect it \cite{Weber_Wulff_2023, gotoman2025accuracy, liang_bias_gpt, UF-new-paper}.
On the UX side, some social media platforms explicitly label AI-generated content, but these labels are only as good as the method for determining whether something is AI-generated in the first place. 

Thus, users and online communities are increasingly taking matters into their own hands: identifying AI-generated content via comments, creating and following social media accounts that aim to educate users in identifying AI-generated content~\cite{jeremyfindsai_instagram_2026}, and turning to social media groups dedicated to AI content identification. Such groups have sprung up on (for example) Facebook and Reddit, and provide hundreds of thousands of discussions of potentially AI-generated content, offering insights into how people perceive and attempt to detect such content. Since people act on their perceptions of the content they see, it is important to understand how people form those perceptions.

However, to the best of our knowledge, there has been no systematic study of what rationales or strategies are employed by internet users to identify AI-generated content in the wild, which limits efforts towards user education and platform design for content labeling.
In this work, we study two communities on Reddit 
to investigate the strategies used to assess content --- and we do so longitudinally over the lifetime of these communities (2 years and 8 months). We investigate the following research questions:
\begin{enumerate}[leftmargin=*,label=\textbf{RQ\arabic*},topsep=3pt]    
\item 
What strategies are used to identify AI-generated content within two prominent subreddits (r/RealOrAI and r/isthisAI)? \label{rq:strategies}
    \item How do strategies change or vary in different contexts, e.g., over time or across different types of content?
    \label{rq:strategies-over-time}
\end{enumerate}

To answer these questions, we collect 13,098 posts and 222,060 comments from two large subreddits, r/RealOrAI (established in 2022, with 1,700 weekly contributions) and r/isthisAI (established in May 2023, with 22,000 weekly contributions), and create a taxonomy of strategies people use to detect AI-generated content by qualitatively analyzing the comments using a combination of human and LLM annotations. We identify and describe the strategies employed by social media users in the wild, and characterize the prevalence of different strategies over time and across different types of content; we do not directly evaluate the effectiveness of the strategies themselves. 

Our contributions include the taxonomy of strategies, as well as qualitative and quantitative observations gleaned from our data. For instance, while we find that people most often use physical cues, such as hands or teeth to detect AI-generated content, we also find evidence to suggest that people are increasingly moving away from relying on fine-grained details, and instead using mental models developed over time about model capabilities and AI trends. 

In summary, our work provides a systematic documentation of the strategies and rationales used by members of two prominent online subcommunities whose mission is to help others identify AI-generated content (or give confidence that something was not AI-generated). 
By systematically cataloging users' AI-detection strategies, we lay the groundwork for user-facing guidance and future research.

%% file: sections/20_related.tex
\section{Related Work}

\header{Perceptions of AI-Generated Content}
Researchers have documented that people's perceptions of content generated using AI is often more negative than their view of content created by other humans. For instance, academics at the University of Florida found that people reported disliking stories they thought were written by AI, even if other people actually wrote them~\cite{uf_chu}. Another study found that people credited artists who used generative AI in competitions with creativity and effort, but did not believe that it required skill~\cite{Lima_2025}. And, a recent Pew report found that most people report being concerned about the growing use of AI in daily life~\cite{kennedy2025americans}. Amongst other findings, the Pew report highlighted people’s concerns that they would not be able to distinguish what was, and was not generated by AI. And these concerns seem well founded: Microsoft reported that people were not skilled at distinguishing AI generated images from “real” images when comparing them side-by-side, particularly when it came to evaluating natural and urban landscapes~\cite{roca2025goodhumansdetectingaigenerated}, or Cooke et al. who find that people's ability to detect AI generated content is no better than a coin toss~\cite{cooke2025good}. While our work does not directly investigate people's perceptions of synthetic, AI generated content to ``organic'' content, we also observed comments that supported both the sentiment that synthetic content was inferior, as well as users who expressed concerns about their ability to distinguish AI-generated content as models progressed in capabilities. 

\header{Automated Detection of AI-Generated Content}
As AI use has grown  
so have automated tools to detect AI-generated content. This is particularly true in the context of detecting AI use in academic settings: several startups market themselves as plagiarism checkers, including 
GPTZero, Originality.ai, Copyleaks, and DetectGPT.

However, there is also skepticism about the accuracy and robustness of AI detection tools. For instance, researchers have found that there are easy ways to evade some tools, from using a single space (token mutation cases)~\cite{cai2023evadechatgptdetectorssingle}, paraphrasing generated content~\cite{krishna2023paraphrasingevadesdetectorsaigenerated, sadasivan2025aigeneratedtextreliablydetected}, and using LLMs themselves through prompts to intentionally evade detectors including prompting imperfections in the output created by generative AI~\cite{lu2024largelanguagemodelsguided}. Detection tools have been found to be unreliable~\cite{UF-new-paper, ernst2025you}, including exhibiting biases or systematic errors. 
For example, content from non-native English speakers is more often falsely identified as AI-generated than content from native English speakers~\cite{Weber_Wulff_2023, gotoman2025accuracy, liang_bias_gpt}.
Further, detection tool performance is brittle to model updates, such as GPT 3.5 to 4~\cite{elkhatat2023evaluating}. Researchers have also found that some methods proposed to make generated content detectable, such as watermarks, can easily be removed using generative AI~\cite{zhao2024invisibleimagewatermarksprovably}. 

In our results, we find limited use of AI detection tools (in part due to subreddit rules), and find that users also rely on a wide range of other, more ad hoc strategies.

\header{Human Detection of AI-Generated Content}
There has also been work that examines people's strategies and abilities to distinguish AI-generated content in different contexts, such as on LinkedIn profile pictures~\cite{mink2022Deepphish, mink_linkedin}, images of faces~\cite{bray2023testing}, AI generated videos~\cite{fu2025learninghumanperceivedfakenessaigenerated}, audio clips~\cite{mai2023warning, diel2024human, muller2022human}, deepfakes~\cite{farid2022creating}. Our work similarly examines people's mental models and reasoning they use to distinguish AI generated content, but does so from the perspective of posts and comments from two subreddit communities, which means that the content that people discuss may overlap with the types of media discussed in prior work. However, because Reddit users can post about anything, our work provides a broader look into a variety of content.

Finally, users may be directly informed that content is AI-generated via platform labels on content; recent work has found that people tend to over-rely on these signals~\cite{holtervennhoff2026s}.

%% file: sections/30_methods.tex
\section{Methods}
To understand strategies people use to identify AI-generated content at scale and over time,  we performed a large-scale mixed-method analysis of 222,060 crowd-sourced, free-text responses on whether 13,098 pieces of media were, or were not, made with AI. 
We discuss the ethical considerations in the Ethics and Adverse Impacts Appencix.

\header{Subreddits of Focus}
To gain a high-resolution understanding of how human identification of AI occurs today and has changed in recent years, we leverage public posts and responses made in two subreddits: `r/RealOrAI', and `r/isthisAI'.
Both of these subreddits exist to provide crowd-sourced answers to whether posted digital media (often photos, videos, or even audio clips), is or is not made with AI. 
While other subreddits exist (e.g., r/AIorFake~\cite{reddit_aiorfake}, r/AI\_or\_Real~\cite{reddit_AI_or_Real}), 
we focus on r/isthisAI  and r/RealOrAI as they are substantially larger and more active, 
with 29,000 and 1,700 average weekly contributions (including posts and comments), respectively. We collect our data from both subreddits using ArcticShift \cite{heitmann_arctic_shift}, a widely available tool that is intended to make Reddit data available to researchers.

\header{Data Collection and Filtering}
We collect every post and comment from r/isthisAI and r/RealOrAI from the communities' creation dates (August 2022 and May 2023, respectively) through March 2026, encompassing 2.8 years of data.

\textit{Posts.} 
To ensure our analyzed posts center on human responses to whether the posted media is or is not AI, we filter our collected posts that: 
(1) Focus on soliciting crowd wisdom; while all posts from r/isthisAI are requests for crowd-sourced human strategies for determining AI, only the posts labeled ``[HELP]'' in r/RealOrAI have this focus, and thus are the only sub-set of posts we collect,\footnote{Other tags, such as ``[GUESS]'' ask users to respond to a user-made challenge, but are not requests for helping determine the authenticity of content by the requester. We also manually identified 136 posts on r/RealOrAI that \textit{would} have been ``[HELP]'' posts (i.e., poster asked for help and did not provide an answer) but appeared before this guideline was enforced in late 2024.}
(2) Are valid; in particular, we remove all posts that have been deleted by their the author or a community moderator,
(3) Are relevant; lastly, we go through all posts, and any that do not focus on crowd-sourced responses to media are labeled as ``irrelevant'' and removed from the dataset and subsequent analysis. 
In total, we analyze 13,098 posts across the two communities (r/RealOrAI: 36.6\%, r/isthisAI: 63.4\%). 

\textit{Comments.}
From our filtered posts, we then collect all user-submitted comments and specifically analyze the subset of them that are:
(1) Direct comments to the post; while comments made generally on the post may be responding to other users' comments, often on tangentially related topics, direct comments are often in response to the post-purposes and thus are a crowd-sourced response to whether the media is or is not AI.
(2) Are valid; we remove all comments that have been deleted, removed, or authored by users who have deleted accounts. 
(3) Are relevant; like posts, we label and remove  comments coded as ``irrelevant''. 
In total, we analyze 222,060 comments (r/RealOrAI: 46.6\%, r/isthisAI: 53.4\%).

\header{Qualitative Analysis: Human Coding}
To characterize the strategies humans use to detect AI, we performed a hybrid qualitative coding methodology~\cite{azungah2018qualitative} to taxonomize these strategies and measure the prevalence across Reddit.
To develop an initial codebook, a single coder began with specific codes selected from Fu et al.~\cite{fu2025learninghumanperceivedfakenessaigenerated} and the SIFT method~\cite{caulfield2019sift}, a digital guide designed to help people establish truthfulness in what they observe; using a randomly selected set of 500, the single coder than inductively added new codes, such as CONTENT\_PLAUSIBILITY, AI\_STYLE, and MODEL\_CAPABILITIES, among others.

Once the initial codebook reached conceptual saturation and no new codes emerged, it was given to a second coder to calculate inter-rater reliability.
One hundred randomly selected comments were independently coded by two coders, and the inter-rater reliability (IRR) of each sub-code was then calculated using Cohen's $\kappa$~\cite{cohen1960coefficient}.
After coding, the coders met to calculate the codebook's IRR, and resolve disagreements; if any sub-code did not achieve a high inter-rater reliability ($\kappa$ $>$ 0.7), this process was repeated.
After three rounds, the IRR for all subcodes was achieved after 300 comments in total. We present the final codebook and IRR values in the Appendix.
Once this codebook was established and verified for reliability, we manually coded 2,000 randomly selected comments and used the resulting human-coded set as ground truth to design and verify our LLM-assisted coding methodology.

\header{Qualitative Analysis: Automated Labeling}
To label all 222,060 posts and comments based on our taxonomy, we developed an automated labeling method and validated it on held-out data.
We used a zero-shot classification pipeline built around Claude Sonnet 4.6~\cite{anthropic2026sonnet46}, structuring each of our codes as an independent classification task.
Using 1,000 manually-coded and randomly sampled comments to serve as a development set,
we iteratively refined our prompt until we achieved high inter-rater reliability between the automated system and the ground-truth human labels, with Cohen's $\kappa > 0.7$~\cite{kraemer2012}.
To validate performance, we then ran our system on an additional \textit{holdout} set of 1,000 human-annotated comments that the automated system had not seen during prompt refinement.
Within this set, we found that every subcode still had substantial agreement ($\kappa > 0.84$ for every subcode).
We then use this AI system to code our full dataset of posts and comments ($N$=222,060), which we analyze and present in our results. We provide the final prompt, along with additional details on the process of creating it, in the Appendix.

\header{Quantitative Analysis: Frequent Words}
To further characterize strategies,
we explore the important words that commonly appear. 
We do this using tf-idf~\cite{salton}, treating the comments within each sub-code as a ``document''. Prior to running tf-idf, we do not stem words (i.e., reduce to roots), and we remove non a-z, 0-9 characters (e.g., removing punctuation). We ignore terms that appear in more than 90\% of documents, and identify the top 10 most unique words for each sub-code.

\header{Quantitative Analysis: Temporal}
Given our full set of labeled comments, we also conducted several quantitative analyses towards answering RQ2, i.e., comparing the frequency of different strategies in comments over time.

Our dataset covers about 33 months, from July 2023 through March 2026. To observe temporal trends, we consider both monthly rolling averages  as well as three 11-month segments of our data. These are:
    \textit{Period 1:} July 7, 2023 -- May 31, 2024; 
    \textit{Period 2:} June 1, 2024 -- April 30, 2025; and 
    \textit{Period 3:} May 1, 2025 -- March 31, 2026.

In both analyses, we compare not the raw number of times a strategy appears (since overall comment volume varies over time) but normalize to the total number of strategies mentioned in each period.
That is, we calculate the {relative proportion of that strategy among all strategies mentioned during that time period}. We normalize by the number of strategies mentioned, rather than the number of comments, since a single comment can contain multiple strategies.

We complement visualizations with statistical tests. We first run $\chi^2$ tests to compare the distribution of strategies observed between periods, correcting for multiple comparisons with Bonferroni. 
To more closely examine how individual strategy proportions change over time, we also run post-hoc tests, again with  Bonferroni corrections.

\header{Limitations}
Our work has limitations similar to prior work that also analyzes general perceptions via Reddit~\cite{bouma2024honestly,oak2025victims,bouma2025scam,wei2024RedditIBSA, nath2026likehammerbuildbreak}.
\textit{First}, while we analyze users' AI detection strategies across two of the most popular crowd-sourced AI-detection communities, it is unclear if the perceptions and strategies here generalize to the general population, who may be less aware, less interested, and interact less with shared opinions for how to detect AI effectively. 
To mitigate, we do our best to contextualize our findings for these communities, while also arguing that the trends within this active group of users continue to provide valuable signals about how beliefs and perceptions of AI have changed over time.
\textit{Second}, while we leverage ArcticShift~\cite{heitmann_arctic_shift} to acquire data, we are also beholden to their data scraping methods and preparation; in particular, this may not reflect the current version of Reddit in which data is later deleted, and also presents comments as tied to the date of the post's creation. To mitigate this, we do our best to remove data and comments that have since been deleted on the live site, and while we cannot track comments to a date beyond the post, we also argue that as active subreddits with thousands of new posts a day, the vast majority of comments likely occur within a few days of the post's creation.
\textit{Third}, while our analysis provides insights into how posted strategies have changed, we ultimately do not claim to identify the root causes of these changes. While we provide hypotheses based on users' self-reported reasoning in posts and technological changes around that time, we ultimately leave this investigation for future work to confirm.
\textit{Fourth}, 
We acknowledge that there are many ways to analyze this dataset, 
such as how media-specific or user-specific characteristics affect posted comments.
We encourage future researchers to leverage our dataset or approach to explore these questions from different perspectives.
\textit{Fifth}, it is reasonable to question whether our AI-assisted methodology may result in biased data that differs from human coding.
To the best of our ability, we mitigate this by comparing our AI-assisted coding to a ground truth of human-coded data made after our codebook's IRR was established.
Thus, we argue that our AI-assisted codings are, by definition, as empirically reliable as any other IRR-verified coding methodology.

%% file: sections/40_results.tex
\input{tables/taxonomy_overview}

\section{Results}

\header{Overview of Dataset}
Our dataset comprises 222,060 comments across r/RealOrAI (103,480, 46.6\%) and r/isthisAI (118,580, 53.4\%) between July 2023 to March 2026. Much of the data comes from the end of 2025 and beginning of 2026 as both subreddits grew in activity. (Additional data on post and comment counts over time is in the Appendix.)

\input{sections/43_qualitative_taxonomy}

\input{sections/44_temporal_taxonomy}
\input{sections/450_case_studies}

%% file: tables/taxonomy_overview.tex
\begin{table*}[t]
\footnotesize \centering

\begin{tabular}{@{}llrl@{}}
\toprule
\textbf{Category or Sub-code}                           & \textbf{Count} & \multicolumn{1}{l}{\textbf{\%}} & \textbf{Description}                                                                      \\ \midrule
\textbf{Perceptions of World}                           &                & \multicolumn{1}{l}{}                     &                                                                                           \\
\hspace{3mm}PHYSICAL\_DETAILS          & 119,390        & 44.5\%                                   & Examines fine-grained details in media                                                    \\
\hspace{3mm}CONTENT\_PLAUSIBILITY      & 66,009         & 16.8\%                                   & Compares to personal experiences and real world        \\ \midrule
\textbf{Perceptions of AI}                              &                & \multicolumn{1}{l}{}                     &                                                                                           \\
\hspace{3mm}AI\_STYLE                  & 36,039         & 9.0\%                                    & Assumptions about characteristics of AI content (e.g., texture)   \\
\hspace{3mm}MODEL\_CAPABILITIES        & 15,597         & 3.0\%                                    & Assumptions about what models are and are not able to do                                  \\
\hspace{3mm}MODEL\_PROMPT\_SPECULATION & 7,653          & 1.5\%                                    & Hypotheses about what models or prompts were used                                         \\ \midrule
\textbf{External Signals}                               &                & \multicolumn{1}{l}{}                     &                                                                                           \\
\hspace{3mm}TRIANGULATING\_INFO        & 21,194         & 5.3\%                                    & Identifies original poster or searches for corroborating evidence 
\\
\hspace{3mm}MEDIA\_AGE                 & 7,191          & 2.1\%                                    & Compares date of content with model capabilities at that time      \\
\hspace{3mm}TOOL\_SIGNALS              & 4,445          & 1.4\%                                    & Use of tools including reverse image search, watermarking, etc.                           \\
\hspace{3mm}OTHER\_EXTERNAL            & 1,385          & 0.2\%                                    & Use of other external sources                                                             \\ \midrule
\textbf{Perceived Motivations for AIG Content}           &                & \multicolumn{1}{l}{}                     &                                                                                           \\
\hspace{3mm}AI\_TRENDS                 & 6,957          & 1.8\%                                    & Uses cues from current events, pop culture, other viral content         \\
\hspace{3mm}MOTIVATION\_FOCUSED        & 6,818          & 1.6\%                                    & Speculates about reasons others might create/spread AI content           \\ \midrule
\textbf{Intuition}                                      & 28,195         & 12.7\%                                   & Short, decisive comments without support or reasons                       \\ \bottomrule
\end{tabular}
\caption{Taxonomy of AI detection strategies in our dataset, including the number of comments in which each strategy appeared, and the percentage this represents of all comments (N=222,060). Note that a single comment may describe multiple strategies. 
}
\label{tab:taxonomy_overview}
\end{table*}

%% file: sections/43_qualitative_taxonomy.tex
\section{\ref{rq:strategies}: Strategies to Identify AI-Generated Content}
\label{sec:qual_analysis_taxonomy}
As shown in Table~\ref{tab:taxonomy_overview}, we find a range of strategies that are used by users to determine whether a piece of media is made with AI. While we find these strategies leverage one of three mental models~\cite{mink_linkedin} for how AI media looks (\textbf{Perceptions of AI}) or what the real world \textit{doesn't} look like (\textbf{Perceptions of World}), we also find heavy use of \textbf{External Signals}, and unspoken \textbf{Intuition}.
We now describe each of these individual codes by their general strategies.

\subsection{Perceptions of World}
To determine whether media was or was not AI, it was often contrasted with what people experience in the real world, and how they differ, or align.
We find that people often focus on what is (not) physically possible, and whether the phenomena captured appear (un)known or (un)common.

\input{sections/43_qual_subsections/431_physical_details}

\input{sections/43_qual_subsections/432_content_plausibility}

\input{tables/casestudies_tfidf}

\subsection{Perceptions of AI}
Often, people determine what is AI based on their pre-conceived notions of what AI. In particular, users often focused on what AI content tends to look like, what they believe the limits of AI capabilities are, and what content tends to be generated from common prompts.

\input{sections/43_qual_subsections/434_ai_style}
\input{sections/43_qual_subsections/435_model_capabilities}

\input{sections/43_qual_subsections/436_model_prompt_speculation}

\subsection{External Signals}
Beyond content itself, users also leveraged external sources, including  understanding where the media originated, when it was created, and how external tools classify it. 

\input{sections/43_qual_subsections/440_triangulating_information}

\input{sections/43_qual_subsections/441_media_age}

\input{sections/43_qual_subsections/439_tool_signals}

\input{sections/43_qual_subsections/442_other_external}

\subsection{Perceived Motivations for AIG Content}

\input{sections/43_qual_subsections/437_ai_trends}
\input{sections/43_qual_subsections/438_motivation_focused}

\subsection{Intuition}

\input{sections/43_qual_subsections/433_intuition}

%% file: sections/43_qual_subsections/431_physical_details.tex
\header{{Physical Details}}
44.5\% of  comments examined detailed features within content, 
whether those were `aligned' with physical reality or physically impossible.
The important words from tf-idf (see Appendix) include specific objects being scrutinized (e.g., \textit{handles}, \textit{tiles}) as well as physical features (e.g., \textit{pupils, elbow} and \textit{index} (finger)).

As people and animals were often the subject of generation, many comments focused on abnormalities within them, such as their hands, fingers, eyes, teeth, whiskers, paws, and fur. For instance, in a post where the original poster was looking for help analyzing the prints on two T-shirts they bought, one comment highlighted how the hands depicted in the image were incorrect, saying:~\pquote{...they are both AI... the first one is the hand holding the cat, a finger going the wrong way}~(User 71185). 
People also examine the {texture and pattern} of elements in the content. They might bring up how someone's skin is too shiny, how the texture of clothing looks incorrect, or how the content appears overall distorted or `melty'. For instance, when evaluating art, one person said: ``\textit{the eyes look melty and classically AI''}
(User 9798).

The disruption of universal physical laws, such as inconsistent lighting, shadows, reflections, or focus, was also a clear indicator used by people. For instance, considering whether promotional content from an electronics manufacturer used AI, part of someone’s comment focused on the shadows that ought to have been cast, saying:~\pquote{Also, the light to the right side of the desk should cause a double shadow on the mini PC. The light from the window seems to be the only light actually casting shadows. The light to the right of the desk is there but it doesn't look like anything is interacting with that light}~(User 52940). 

For videos, consistency of these factors over time was a common cue. Odd motion, or how objects change over time, often caused suspicion: \pquote{The faces are so so wrong looking, they don't move right, this is definitely AI}~(User 67645).

While these details are often used to justify that something is AI-generated, a few were used to rationalize authentic content. For instance, the realistic motion and the consistent lettering in the video clip of a child on rollerskates was used by one person to justify that the media didn't involve AI: \pquote{[I am] Leaning real. [The] Letters on the shops don't change or morph}~(User 9632).

%% file: sections/43_qual_subsections/432_content_plausibility.tex
\header{{Content Plausibility}}
16.8\% of posted comments broadly assess whether the content {aligns with their expectations of how people, animals, and the world behave.}
Unlike PHYSICAL\_DETAILS which tends to focus on minor, but clear inconsistencies reminiscent of AI creation, comments here focus on general behavior trends that are common or uncommon.
For instance, when discussing a video clip showing baby and a cat interacting, one user comments:~\pquote{Babies do not nearly have that much coordination to both be standing up, talking so well and holding up that large cat. It is obvioulsy [sic] AI} (User 12140). 
Similarly, another person comments on a clip of horses, saying, \pquote{I don't see anything screaming fake. The horses are looking at/reacting to each other realistically.
}~(User 4581).

%% file: tables/casestudies_tfidf.tex
\begin{table*}[tb]

\small \centering
\begin{tabular}{@{}llllll@{}}
\toprule
AI STYLE  & AI TRENDS  & MEDIA AGE & \begin{tabular}[c]{@{}l@{}}MODEL / PROMPT\\ SPECULATION\end{tabular} & \begin{tabular}[c]{@{}l@{}}MOTIVATION\\ FOCUSED\end{tabular} & \begin{tabular}[c]{@{}l@{}}TOOL\\ SIGNALS\end{tabular} \\ \midrule
glossy    & popping    & preai     & diffusion                                                            & butchering                                                   & synthid*                                               \\
\rowcolor[HTML]{EFEFEF} 
soulless  & flooded    & 2016      & dalle                                                                & crypto                                                       & wwwtorfai                                              \\
sheen     & doorbell   & spaghetti & grok                                                                 & ragebait                                                     & tineye                                                 \\
\rowcolor[HTML]{EFEFEF} 
dashes    & spaghetti  & covid     & generates                                                            & waters                                                       & isthisaicom                                            \\
outlines  & chihuahua  & 2014      & chatgpts                                                             & scammy                                                       & sightengine                                            \\
\rowcolor[HTML]{EFEFEF} 
flour     & trends     & january   & typed                                                                & victim                                                       & aidetective*                                           \\
yellowish & sleeping   & dated     & flux                                                                 & goal                                                         & watermarked                                            \\
\rowcolor[HTML]{EFEFEF} 
cartoony  & trampoline & predate   & input                                                                & gain                                                         & cybyapi                                                \\
brushes   & porch      & 2013      & prompter                                                             & sympathy                                                     & assigned                                               \\
\rowcolor[HTML]{EFEFEF} 
hue       & react      & october   & brushes                                                              & motivation                                                   & analyzed                                               \\ \bottomrule
\end{tabular}
\caption{Top 10 unique keywords in specific subcodes (resulting from tf-idf). Note that punctuation has been removed: we shortened some urls in TOOL SIGNALS and represent this with an asterisk. Words are ordered in decreasing tf-idf value.}
\label{tab:tfidf_casestudies}
\end{table*}

%% file: sections/43_qual_subsections/434_ai_style.tex
\header{{AI Style}}
9\% of posted comments consider whether the content looks like or includes tell-tale signs of common AI-generated outputs. 
For instance, one user began their comment by saying \textit{``Several classic hallmarks of AI image generation are evident...}'' (User 33021). 
These tells vary by the media type being evaluated.
When considering text, people often noted that the presence of em dashes or the use of particular sentence structures (e.g., examples in groups of three, generic or overly ornate but meaningless phrases) was indicative of AI-generated content. Comments on images often consider specific font choices in images, and while comments on audio often look for `generic' speech. In video, this is often seen in overly smooth motion, emotionless eyes, and specific events such as yelling in the first few seconds or the movement of liquids, such as water, which are common in AI-generated videos. For instance: ``\textit{...If you've seen other AI videos, you know, that liquid motion is one of the hardest things for it to emulate naturally}'' (User 56774). %

Table~\ref{tab:tfidf_casestudies} lists the words commonly (uniquely) used to describe AI style, including words describing the texture of content (e.g., \textit{glossy} and \textit{sheen}). 
There are also specific markers that people associate with AI content (e.g., \textit{dash} as in em-dash), and specific color palettes (e.g., \textit{yellowish}, \textit{hue}). For instance: ``\textit{It’s not just misspelling but the em dash, and the kerning. A human didn’t write that}'' (User 74675). 

Some people noted limitations of this approach, observing that because models are trained on data from existing artists, their outputs could mimic the styles of those artists --- making the original artists' work appear, now, to be (incorrectly) stylistically AI. 
For instance, 
someone said, ``\textit{...you’re also making it harder for real artists. Some of you are convincing people that *actual* artistic quirks, mistakes, and techniques are `obvious' AI. AI was partially trained on real, stolen, art''}
(User 63442). 

%% file: sections/43_qual_subsections/435_model_capabilities.tex
\header{{Model Capabilities}}
3\% of posted comments leverage the users' perception of what AI systems can, and cannot, do.
For instance, people might assume that AI is not capable of creating realistic water movement, and use that assumption as the basis for believing that all beachside images must therefore be authentic. Another common example was assumptions around models' (un)abilities to produce legible, well-kerned text in images. Other examples connect with the details that users identify under the PHYSICAL\_DETAILS and AI\_STYLE sub-codes: e.g., models' inability to keep moving objects consistent, to represent physics realistically, or to use certain art styles in appropriate contexts.
For instance, one user opining on a video wrote, \textit{``AI still doesn't seem to understand eye movements somehow''} (User 18658). %
 
People also considered how models could outperform humans. For example, after identifying the original account responsible for YouTube videos, people used the fact that this account frequently posted long (over 30 minute) videos to determine that it must be AI-generated, as only AI could produce that volume of content so quickly.

Interestingly, we find that these beliefs are not only related to inherent model capabilities but also to the quirks of particular AI systems and side channels they may have. 
For instance,  AI systems such as Sora 2 allowed users to generate clips of up to 15 seconds of high-quality, realistic video clips, even if they did not pay for the service~\cite{peckham2025sora2}.
As these systems were both commonly used and posed a limit to what lay-AI creators could generate, people often used them as indicators to question whether AI was used, as in this case with User 112525 who said \pquote{Well considering it's longer than 15 seconds it probably not AI}, as to confirm AI's presence: \pquote{[the] Video is also exactly 10 seconds long.. not looking good} (User 74212).

%% file: sections/43_qual_subsections/436_model_prompt_speculation.tex
\header{{Model and Prompt Speculation}}
Another strategy people use to support their assessment that something is AI-generated is to speculate about the models or prompts that were used to generate the content in question (the MODEL\_PROMPT\_SPECULATION sub-code).
For example, people often {referred to specific AI companies} when identifying elements in content that tipped them off to believe it was AI-generated:
e.g., ``ChatGPT-style'' lettering on a bottle (User 5428),
word choice or images typical to ChatGPT,
or a ``Sora accent'' in how people speak in videos (User 7941).
Sometimes, people even identify specific model versions, 
such as ``4o'' as shorthand for OpenAI's GPT-4o. 

Considering the tf-idf results in Table~\ref{tab:tfidf_casestudies}, common words under this sub-code include references to AI or prompting techniques (e.g., \textit{diffusion}, \textit{prompter}), as well as references to specific companies and models (e.g., \textit{dalle}). 

Moreover, people sometimes developed guesses about the prompt that could have been used to generate the content. For example, people sometimes identified prompts that they had previously seen produce similar types of content --- e.g., observing that the content looks similar to ChatGPT output when it is prompted for ``voodoo,'' ``magic,'' or ``occult'' (User 4266).
Sometimes users experimented directly by attempting to recreate similar output with specific models.

People also mused that if they \textit{could not} imagine a prompt that could have produced the content, that was a signal that content was not AI-generated. For example, User 39411 discussed how they would not know how to prompt an AI model to create a video clip that so accurately reflected real-world physics, thus leading them to a ``not AI'' conclusion.

%% file: sections/43_qual_subsections/440_triangulating_information.tex
\header{{Triangulating Information}}
5.3\% of posted comments triangulate using information from external trustworthy sources (e.g., ``lateral reading''~\cite{lateralreading}), and using either it's presence or absence, as a signal of whether content is likely AI-generated.

Some users attempt to identify and evaluate the original source, such as specific social media accounts or news outlets. 
Sometimes, the account that produced the content discloses the use of generative AI; other times, other content on the profile dicloses or is interpreted to involve AI use. 

Users may also find corroborating evidence from third-party sources via search engines (e.g., Google) , news sources (e.g. CNN), or fact-checking services (e.g., BBC Verify). For example, \textit{bbc} appears in the tf-idf results for this strategy (see Appendix). 
Or, when evaluating a video, one user wrote \textit{``It's real and has been verified by CNN and other news outlets''} (User 45272).

%% file: sections/43_qual_subsections/441_media_age.tex
\header{{Media Age}}
2.1\% of posted comments use the {age of the content} in their decision-making process.
Often, this involved determining which AI models were available by that point, and whether during that period such content could be created (see \textbf{Model Capabilities}).
In some cases, the age of content rules out AI entirely when images were decades old; in other cases, users combine content age with their mental models about model capabilities at that time. For example: ``pic is almost 4 years old... anything fine-looking older than 2023 is automatically real'' (User 58056).
This age was established via a number of methods, including reverse image search, 
and their existing knowledge of content or events, and the timeframe they occurred within.

Beyond straightforward keywords referring to time (e.g., \textit{dated} or \textit{2013}), Table~\ref{tab:tfidf_casestudies} shows the prevalence of \textit{covid} as a time reference.
For instance: 
 \textit{``Not AI, I bought this exact art pre covid for my old place''} (User 107915).

%% file: sections/43_qual_subsections/439_tool_signals.tex
\header{{Tool Signals}}
1.4\% of comments use AI detection tools, watermarks, or ad hoc tools. Specific tool names that appear commonly (Table~\ref{tab:tfidf_casestudies}) include \textit{synthid} (watermarking), \textit{tineye} (reverse image search), and several third-party AI detection tools (e.g., \textit{torf.ai}, \textit{aidetective}). 
We also observed examples of users attempting to use AI chatbots themselves as AI detectors, without any embedded tools.

The relatively limited prevalence of this strategy in our dataset is influenced by the community rules of r/isthisAI:
``AI detectors are unreliable and should not be used as evidence for AI or not.''. However, r/RealOrAI does not have a similar rule, and r/isthisAI does allow use of watermarks: ``SynthID is not an AI detector and therefore allowed.''

In some cases, people look for watermarks manually: \textit{``You can see the blurred out sora logo''} (User 11913).
In other cases, people look for cryptographic watermarks, such as SynthID: \textit{``If you run this through SynthID it will detect that it's made using nano banana''} (User 16526).
Unfortunately, even when people knew about cryptographic watermarks, they sometimes misunderstood them.
For instance, one user incorrectly expected a watermark to be detectable by multiple tools:  \textit{``I used chat GPT, I used Grok and I even used Deep. Those assistants were not able to tell even after scanning the image. Ford me that's a little inconclusive even though Google confirmed it with the SynthID''} (User 21684).

Lastly, several users reported ad hoc strategies that repurposed existing tools. For example, people often use reverse image search to aid in other strategies: ``Reverse image search tells me \pquote{gigantic rat in rice fields' seems to be a bit of an AI video trend} (User 23545). In one unique example, a commenter evaluated whether a prematurely leaked movie trailer was AI-generated or genuine by uploading the video to YouTube and relying on YouTube's detection of copyrighted content to determine that it was not AI-generated.

%% file: sections/43_qual_subsections/442_other_external.tex
\header{Other External}
Finally, 0.2\% of posted comments relied on sources or knowledge not embedded in the media itself, but not clearly supported by common information infrastructures. For instance, we observed people using content in other comments and other general forum activity as signals in their decision making, such as this individual who noted that the photographer of an animal provided context for the image, saying \textit{``Yeah the photographer who took it mentions it has an injured back leg''} (User 240). The tf-idf results also surfaced \textit{youtube} as a common word in this category.

%% file: sections/43_qual_subsections/437_ai_trends.tex
\header{{AI Trends}}
1.8\% of posted comments consider whether the posted content follows trends and patterns in previously-identified and often viral AI-generated media. 
For example, AI-generated videos of home surveillance cameras (e.g., Ring doorbells) became popular in 2025~\cite{fauxsurveillance}, and this made User 7665 believe that all such videos were likely AI: \pquote{Ring/security camera type of video? Check... All of these traits scream AI generated video right away.} Other common trends mentioned were dashcam videos, Halloween videos, and Ghibli-style pictures.
For example, one user remarked sarcastically: \textit{``Boy, I'm sure glad all these brand new videos of animals being spooked by Halloween decor next to a full bowl of candy are coming out right BEFORE Halloween''} (User 13547).

Common words describing AI\_TRENDS (Table~\ref{tab:tfidf_casestudies}) fall into two categories. First, those that express the sheer volume of certain types of content: \textit{popping} (as in, ``popping up in my feed''), \textit{flooded}, and \textit{trends}). 
And second, references to specific trends (e.g., \textit{doorbell} and \textit{porch} in references to AI-generated doorbell camera videos).

Some users tied these trends to specific platforms and target demographics, e.g.,  Facebook often having AI-generated images of older adults celebrating birthdays alone or of children showing off unrealistically highly-skilled craftwork. For instance: \textit{``Facebook is littered with this slop and everyone’s grandparents believes it to be real''} (User 25062). 

%% file: sections/43_qual_subsections/438_motivation_focused.tex
\header{{Motivation For Content}}
1.6\% of posted comments, people {speculate about what incentives someone might post certain AI-generated content}, and then use those perceived motivations whether the content likely is, or is not, AI. 

Important words here (Table~\ref{tab:tfidf_casestudies}) reveal common motivations like perpetrating scams (e.g., \textit{butchering}, as in ``pig butchering'' scams) 
and engagement bait (e.g., \textit{ragebait} and \textit{sympathy}).
Other motivations discussed include
avoiding paying for the rights of commercial images, 
generating revenue/promote products,
and political misinformation or propaganda. 
For example, in response to a video related to a politically-charged current event, one user posted: \textit{``This one is definitely AI. Be aware that Trump's supporters will pump this out and it will be crucial that we help each other in determining what is real and what is fake'}' (User 20404). 
In another example, one user said, \textit{``I wouldn’t trust any images coming from Iranian propaganda, they have been caught using AI images numerous times already''} (User 105113).

%% file: sections/43_qual_subsections/433_intuition.tex
Lastly, we find that 12.7\% of posted comments are decisive, often short statements without additional reasoning.
Most of the important words in INTUITION  (see tf-idf in the Appendix) suggest confidence: e.g., \textit{fakest}, \textit{painfully}, \textit{christ}, and \textit{1000000}. 
For instance, one person wrote: \textit{``This is the fakest thing ever. 1000000\% AI''} (User 424). People also refer to a different subreddit, r/isthisacirclejerk, to make the point that something is obviously AI: \textit{``I thought this was the r/isthisacirclejerk sub lol. Extremely ai''} (User 16484).

%% file: sections/44_temporal_taxonomy.tex
\section{\ref{rq:strategies-over-time}: How Strategies Vary}

We now investigate: how do people's uses of different AI detection strategies change or vary: over the lifetime of the communities, or across different types of content?

\begin{figure}[t]
    \centering
        \includegraphics[width=\linewidth]{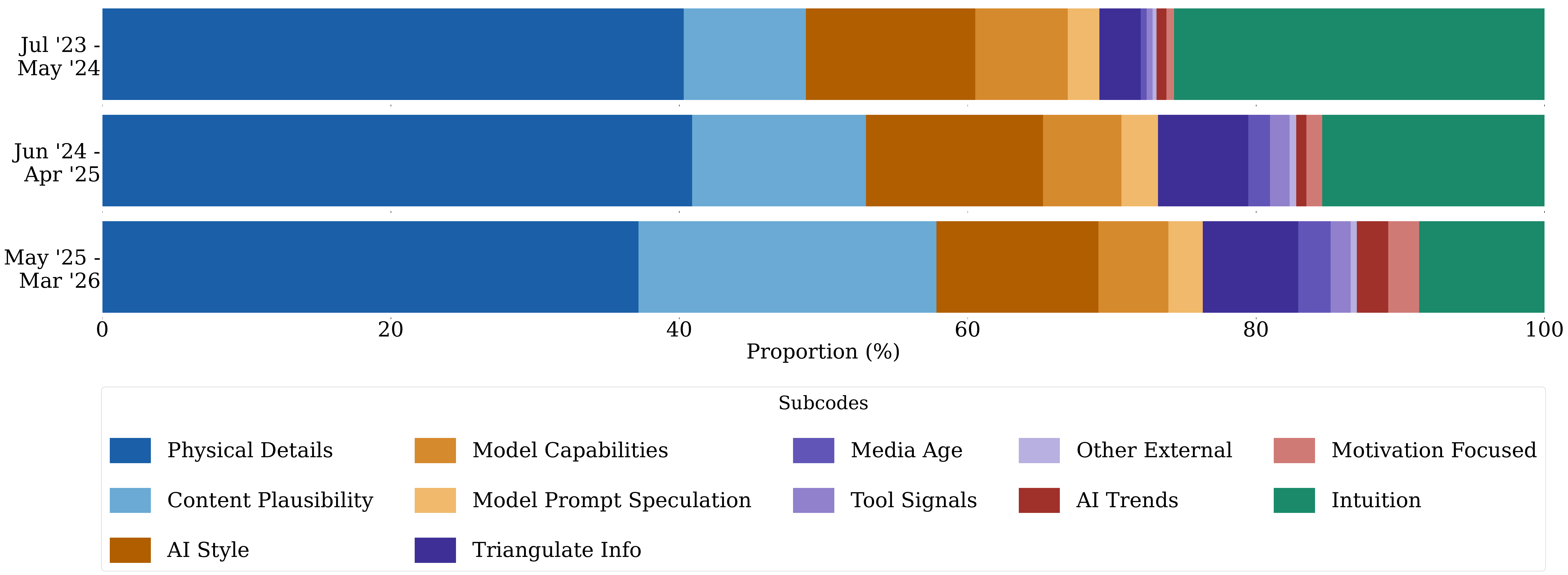}
        \caption{AI Strategies over time: 
        The proportion of each sub-code across our observed time periods. 
        }
        
    \label{fig:subcodes_over_time}
\end{figure}

\input{sections/442_statistical_analysis}

%% file: sections/442_statistical_analysis.tex
\subsection{Changes Over Time}
In Figure \ref{fig:subcodes_over_time}, we display the relative frequencies of different strategies used over three 11 month intervals in our dataset: (\textbf{p1}) July 2023--May 2024, (\textbf{p2}) June 2024--April 2025, and (\textbf{p3}) May 2025--March 2026.
To evaluate whether the relative proportions of strategies reported in comments vary across time, we conduct a $\chi\textsuperscript{2}$ test of independence between our time intervals (p1--p2 and p2--p3); and we apply a Bonferroni correction to control for false positives.

Indeed, we find that we find the proportions of reported AI strategies significantly vary between p1 and p2 ($\chi^2$ ($11$, $N${=}$3{,}165$){=}$65.18$; $p_\text{adj}{<}0.025$)
as well as p2 and p3 ($\chi^2$ ($11$, $N${=}$320{,}141$){=}$267.69$; $p_\text{adj}{<}0.025$)
To evaluate what and how specific strategies change across time, we then conduct a pair-wise series of post-hoc $\chi\textsuperscript{2}$ tests for each AI strategy within the three compared time periods (e.g. p1--p2, and p2--p3), and again apply a Bonferroni correction among each set of post-hoc tests ($p_\text{adj}{=}0.002$). As shown in  Table~\ref{tab:posthoc_subcode}, we find that 7 AI strategies varied significantly in proportions across our evaluated time periods.

\input{tables/posthoc_subcode}

Between p1 (July 2023--May 2024) and p2 (June 2024--April 2025), we find that only two strategies significantly varied in prevalence:
\begin{enumerate}
\item Intuition decreased, perhaps due to stricter moderator enforcement of community guidelines that prohibit short statements without providing support.
\item Triangulate Info increased, perhaps as users gained experience with AI detection or as the type of content posted changed.
\end{enumerate}

Between p2 (June 2024--April 2025) and p3 (May 2025--March 2026), we find that five strategies varied significantly in prevalence:
\begin{enumerate}
\item Intuition decreased: As above, this may be due to changing subreddit norms and rules.
\item Physical Details decreased:  This relative decline may be in part due to the rise of other strategies (discussed next) that develop as subreddit users gain longitudinal experience with AI-generated content and the surrounding online ecosystems. 
\item AI Trends increased: As AI use on social media became more popular and viral successes inspired copycat content, AI trends also appear more often in posts and thus comments.  
For example, in p3, User 6317 mentions \textit{``Someone is trying to copy the success of the bunnies''} from mid-2025~\cite{thomas_ai_2025}.
The Appendix also contains additional longitudinal data for specific trends-related keywords (see ``AI Trends Over Time'').
\item Content Plausibility increased, perhaps in place of comments expressing a non-justified intuition. 
\item Motivation increased: p3 overlaps the 2025 holiday shopping season, during which time there was a growing awareness of AI-generated product scams~\cite{dropshipping}, reflected in our data.
For example, \textit{``this is totally ai. dont use etsy anymore. there's so much dropshipping and ai bullshit on there meant to just make as much money as possible, disappear, then reappear''} (User 17319). 
\end{enumerate}

\subsection{Variation Across Content}

Different strategies may be used for different types of content. Below, we investigate differences between (eventually-concluded) AI versus non-AI content. 
We provide an additional content-based case study (politics) in the Appendix. 

%% file: tables/posthoc_subcode.tex
\begin{table*}[]
\small \centering

\begin{tabular}{@{}l|lll|llr|lrl@{}}
\toprule
                           & \multicolumn{3}{c|}{\textbf{Counts}}    & \multicolumn{3}{c|}{\textbf{p1 vs p2}}                                      & \multicolumn{3}{c}{\textbf{p2 vs p3}}                                   \\ \midrule
\textbf{Labels}            & \textbf{p1} & \textbf{p2} & \textbf{p3} & \textbf{$\chi^2$} & \textbf{Delta}    & \multicolumn{1}{l|}{\textbf{p-val}} & \textbf{$\chi^2$} & \multicolumn{1}{l}{\textbf{Delta}} & \textbf{p-val} \\ \midrule
PHYSICAL\_DETAILS          & 295         & 995         & 118,100     & 0.060             & 1.48\%            & 0.807                               & \textbf{14.170}   & \textbf{-9.10\%}                   & \textbf{0.000} \\
\rowcolor[HTML]{EFEFEF} 
CONTENT\_PLAUSIBILITY      & 62          & 293         & 65,654      & 6.859             & 42.18\%           & 0.009                               & \textbf{109.218}  & \textbf{71.60\%}                   & \textbf{0.000} \\
AI\_STYLE                  & 86          & 299         & 35,654      & 0.108             & 4.60\%            & 0.740                               & 2.652             & -8.68\%                            & 0.103          \\
\rowcolor[HTML]{EFEFEF} 
MODEL\_CAPABILITIES        & 47          & 132         & 15,418      & 0.867             & -15.50\%          & 0.352                               & 1.591             & -10.55\%                           & 0.207          \\
MODEL\_PROMPT\_SPECULATION & 16          & 62          & 7,575       & 0.175             & 16.58\%           & 0.675                               & 0.213             & -6.44\%                            & 0.644          \\
\rowcolor[HTML]{EFEFEF} 
TRIANGULATE\_INFO          & 21          & 152         & 21,021      & \textbf{11.785}   & \textbf{117.77\%} & \textbf{0.001}                      & 0.474             & 5.91\%                             & 0.491          \\
MEDIA\_AGE                 & 3           & 37          & 7,151       & 4.711             & 271.06\%          & 0.030                               & 5.535             & 48.01\%                            & 0.019          \\
\rowcolor[HTML]{EFEFEF} 
TOOL\_SIGNALS              & 3           & 33          & 4,409       & 3.681             & 230.95\%          & 0.550                               & 0.002             & 2.32\%                             & 0.964          \\
OTHER\_EXTERNAL            & 2           & 11          & 1,372       & 0.112             & 65.47\%           & 0.738                               & 0.000             & -4.48\%                            & 1.000          \\
\rowcolor[HTML]{EFEFEF} 
AI\_TRENDS                 & 5           & 17          & 6,935       & 0.000             & 2.29\%            & 1.000                               & \textbf{24.340}   & \textbf{212.40\%}                  & \textbf{0.000} \\
MOTIVATION\_FOCUSED        & 4           & 27          & 6,787       & 1.306             & 103.08\%          & 0.253                               & \textbf{11.725}   & \textbf{92.50\%}                   & \textbf{0.001} \\
\rowcolor[HTML]{EFEFEF} 
INTUITION                  & 188         & 375         & 27,632      & \textbf{39.884}   & \textbf{-39.99\%} & \textbf{0.000}                      & \textbf{135.573}  & \textbf{-43.57\%}                  & \textbf{0.000} \\ \bottomrule
\end{tabular}
\caption{Chi-squared post-hoc results, comparing how strategies change across periods (with comparisons between period 1). Rows in bold font indicate statistically significant results, where the \textit{p-value} is less than 0.002 (due to Bonferroni corrections). 
Delta percentages are based on the proportion of stategy-mentions in a particular period, i.e., the data underlying Figure~\ref{fig:subcodes_over_time}.}
\label{tab:posthoc_subcode}
\end{table*}

%% file: sections/450_case_studies.tex
\input{tables/posthoc_solved}

\header{Strategies for AI vs. Non-AI Concluded Media}
We observed in our qualitative analysis that eleven of the twelve strategies (the exception being AI\_STYLE) were used both to justify decisions that something \textit{was} AI-generated 
as well as that it \textit{was not} AI-generated. However, are there different strategies used more or less often in the two cases? 

In November 2025, moderators on r/isthisAI introduced new flairs (specifically, for posts concerning images or videos) that allowed users or moderators to update a post once it has been established that the content in question is or is not AI generated~\cite{newflairs}. 
Though these labels exist only for a limited number of posts in our dataset --- 843 comments associated with posts that use the ``Solved [AI]'' flair, and 378 comments associated with posts that use the ``Solved [Not AI]'' flair --- they give us an opportunity to compare strategies that users apply in the two cases.

Indeed, we find significant differences.
Via a $\chi^2$ test for independence between posts labeled as ``AI'' vs ``Not AI'', we find that the proportion of AI strategies vary significantly 
($\chi^2$ ($11$, $N${=}$1{,}711$) = $195.80$; $p{<}0.05$).
As shown in Table~\ref{tab:posthoc_solved}, we evaluate which strategies vary in proportion via a series of pairwise, Bonferroni-corrected, $\chi^2$ tests among every strategy and find that two significantly vary ($p_\text{adj}{<}0.002$).

Two strategies are significantly different for AI compared to non-AI content:
\begin{enumerate}
\item AI Style increases (+318.8\%): Users typically use AI style to judge content as AI-generated. AI style comments on ``Solved [not AI]'' posts involve either users making incorrect judgments that something \textit{is} AI-generated, identifying that something is authentic but modified (\textit{``Verdict: Real but edited with AI''}, User 19924),
or identifying non-AI inauthentic content (\textit{``Not AI. Just shitty Photoshop''}, User 70807). 
\item Triangulating Info decreases (-60.5\%): In both cases, people use external evidence to support their decision; we hypothesize that such evidence may be more common or easier to find for non-AI content. 
\end{enumerate}

%% file: tables/posthoc_solved.tex
\begin{table*}[tb]
\small \centering

\begin{tabular}{llllrl}
\hline
\textbf{Sub-code}          & \textbf{AI}  & \textbf{Not AI} & \textbf{$\chi^2$} & \multicolumn{1}{l}{\textbf{Delta}} & \textit{\textbf{p-val}} \\ \hline
PHYSICAL\_DETAILS          & 437          & 199             & 0.533             & -5.2\%                             & 1.000                   \\
\rowcolor[HTML]{EFEFEF} 
CONTENT\_PLAUSIBILITY      & 175          & 95              & 3.569             & -20.5\%                            & 0.707                   \\
\textbf{AI\_STYLE}         & \textbf{194} & \textbf{20}     & \textbf{49.175}   & \textbf{+318.8\%}                  & \textbf{0.000}          \\
\rowcolor[HTML]{EFEFEF} 
MODEL\_CAPABILITIES        & 38           & 14              & 0.131             & +17.2\%                            & 1.000                   \\
MODEL\_PROMPT\_SPECULATION & 39           & 3               & 9.737             & +461.3\%                           & 0.022                   \\
\rowcolor[HTML]{EFEFEF} 
\textbf{TRIANGULATE\_INFO} & \textbf{65}  & \textbf{71}     & \textbf{32.971}   & \textbf{-60.5\%}                   & \textbf{0.000}          \\
MEDIA\_AGE                 & 4            & 47              & 92.922            & -96.3\%                            & 0.000                   \\
\rowcolor[HTML]{EFEFEF} 
TOOL\_SIGNALS              & 29           & 10              & 0.198             & +25.2\%                            & 1.000                   \\
OTHER\_EXTERNAL            & 2            & 2               & 0.103             & -56.8\%                            & 1.000                   \\
\rowcolor[HTML]{EFEFEF} 
AI\_TRENDS                 & 46           & 11              & 2.790             & +80.6\%                            & 1.000                   \\
MOTIVATION\_FOCUSED        & 26           & 8               & 0.438             & +40.3\%                            & 1.000                   \\
\rowcolor[HTML]{EFEFEF} 
INTUITION                  & 140          & 36              & 8.263             & +67.9\%                            & 0.0485                  \\ \hline
\end{tabular}
\caption{Chi-squared post-hoc comparison of strategies used in ``Solved [AI]'' vs ``Solved [Not AI]'' posts. Bolded rows are statistically significant, post-Bonferroni corrections.}
\label{tab:posthoc_solved}
\end{table*}

%% file: sections/500_discussion.tex
\section{Discussion and Conclusion}

Finally, we step back and reflect on (1) lessons about the mental models that we see people developing around AI-generated content detection, and (2) what our findings suggest about potential approaches to support uses in this task.

\input{sections/501_mental_models}
\input{sections/502_solutions}

%% file: sections/501_mental_models.tex
\subsection{Mental Models for Detecting AI Content}

Stepping back from individual AI detection strategies, we draw several higher-level lessons about the mental models that commenters in our dataset have developed.

\header{Content-focused strategies dominate strategies that rely on specific external signals} 
In our dataset, the most commonly used strategies consider only the specific piece of content under evaluation, often combined with a user's evolving mental model of what AI-generated content ``looks like'' or what models are capable for producing. By contrast, strategies that rely on external trustworthy sources or tools are much more rare. This is not necessarily a good or a bad thing (indeed, AI detection tools are not robust, as we have discussed, and appear less in the data because of one of the subreddit's rules). Still, it suggests a potential opportunity for an increased role of authoritative external tools and/or resources; though it also suggests potential challenges for such external resources, since users may prefer the lower-effort, content-focused heuristics.

\header{Many popular strategies are detail-oriented or model-specific, and may not be robust to AI model evolution}
Our findings suggest that people are developing mental models of what AI-generated content looks like, built on personal experiences and beliefs developed over time. People often focus on details to identify AI-generated content, or to justify their intuition --- focusing on physical details (e.g., hands), the plausibility of certain aspects of the content (e.g., a child's behavior), or specific indications of AI style (e.g., water texture). These details are then intersected with the assumptions that people have developed about the capabilities of AI models. While these strategies can be very effective in some cases, they are likely not robust to evolutions in AI models; and indeed, in some ways this is a fundamental ``arms race''. Thus, such detail-oriented strategies should be complemented with others, which rely on external sources and/or a broader understanding of the ecosystem. 

\header{Some heuristic strategies rely on broader context or experience that particularly vulnerable users may not have}
Our results suggest increasing use of strategies that rely not just on the detailed characteristics of a piece of content, but on a broader understanding of the online ecosystem: what motivates people to post AI-generated content, and what types of AI-generated memes are currently going viral. While these may be effective strategies for users who are very ``online'', such as Reddit commenters and some of this paper's authors, they rely on exposure and experience over time that many people may not have (e.g., older adults, or less tech-savvy users). The consequence is that these people may be more vulnerable to AI-generated content, which can be problematic if that content also spreads mis/disinformation (e.g., political or health). Moreover, because these types of strategies rely on accumulated experience across many pieces of content, they may be hard to teach.

\header{Mental models are socially constructed, and there is a risk of false confidence from unreliable tools or strategies}
Particularly in our dataset, where users interact specifically for the purpose of discussing AI-generated content detection, the emerging strategies and underlying mental models are socially constructed and reinforced. The same is true in other contexts where users engage in collective sensemaking, e.g., in the comments on a Facebook post. 

This can be helpful to people, as they learn and spread information about how to recognize AI-generated content. However, this can also have unintended negative effects, such as when the strategies people share are outdated relative to model capabilities, when AI detection tools are used and recommended but not actually effective, or when incorrect strategies may systematically categorize `non-AI' media as `AI' according to held biases of others' identities~\cite{mink_linkedin}. The resulting reliance on such strategies or tools can, in turn, risk creating false confidence that a particular piece or type of content is authentic, or that authentic content is fictitious.

%% file: sections/502_solutions.tex
\subsection{Lessons for Existing Solutions}

We reflect on what our findings suggest about proposed mechanisms for helping users identify AI-generated content.

\header{Watermarking} 
Prior academic work has discussed alternative methods for AI detection, including both watermarking and other cryptographic solutions. These solutions provide several benefits, including clear provenance information and the opportunity to rely on detection tools rather than individual human knowledge. 

However, in addition to other limitations with such approaches (discussed in the Related Work), our work highlights important usability challenges. For instance, we observed people who knew that SynthID (Google's watermarking) indicated generative AI use, but \textbf{misunderstood how to check for it}, e.g., thinking that they could ask another service (such as ChatGPT) to check for the Google watermark.

Thus, we find that it is not enough to develop the technical capabilities the watermarking and cryptographic techniques enable. It is also important to consider the \textbf{usability} of these tools as well, both in the interface that companies present to people, as well as the information they release alongside these tools.

\header{AI Disclosures}
Existing research has demonstrated limitations of AI content disclosures that are starting to be deployed on some platforms, e.g., that users may not fully understand what the disclosures mean~\cite{holtervennhof2025, gamage2025labeling, moller2026impact}. Our findings show that people \textit{are} sometimes leveraging these disclosures, e.g., looking for cues like AI disclosures in an account profile or in an account's other content to assess a piece of content under scrutiny. However, this process is ad hoc and disclosures currently vary widely across platforms. This lack of standardization may muddy waters for users. Rather than reinvent the wheel every time when considering disclosures, \textbf{we urge companies to collaborate and unify disclosures} in a way that prioritizes user understanding --- e.g., via the existing C2PA coalition~\cite{c2pa}. 

\header{Consumer Education}
Though the burden should certainly not fall on users alone to identify AI-generated content, consumer education and guidance can play a role in helping people develop accurate mental models and effective strategies. Our work  highlights the importance of doing \textbf{human-centered research when developing and releasing guidance} to understand the strategies people are already using, which strategies have staying power, which strategies are misunderstood, and how those strategies are succeeding or failing. For example, we find that people often consider the details of specific content and rely on their assumptions of AI model capabilities --- though these strategies can be effective, education for users should also convey that detail-focused strategies can become outdated as models improve. Echoing significant past work in the broader space of online mis/disinformation, we encourage the development of guidance and education that emphasizes strategies that will work even as AI models evolve, and do not disproportionately harm groups of people along the way~\cite{mink_linkedin} e.g., lateral reading, triangulation, and relying on trustworthy external sources~\cite{caulfield2019sift, kampen_library_nodate}.

%% file: sections/600_futurework_limitations.tex
\subsection{Future Work}

With this paper's publication, we will also release our labeled dataset, enabling future research to dig into questions that were outside our scope here. For example, future work might: investigate how strategies vary across media type or content type; dive deeply into political content in particular (labeled with a ``flair'' in r/isthisAI); and investigate the effectiveness of strategies (e.g., by leveraging the ``[GUESS]'' posts on r/RealOrAI that provide a ground-truth answer).

More broadly, future work might complement our investigation with studies of other user populations or datasets; with in-depth interviews to dig deeper into people's AI detection mental models; with analyses of publicly-available guides for identifying AI-generated content; and with investigations of the impacts of AI detection tools and platform labels on people's skills and perceptions.

In summary, our work provides insight into evolving user mental models for AI detection, and we highlight lessons for supporting users in this task. 
In addition to our taxonomy of strategies, we will make our labeled dataset public at the time of publication to support future research.

%% file: sections/800_statements.tex
\input{sections/810_ethics}

%% file: sections/810_ethics.tex
\section{Ethics and Adverse Impact Statement}
\label{sec:ethics}

We describe here several ethical considerations that arose in the design and execution of our research:

\begin{enumerate}

\item Our University's IRB does not consider studies like this one to be federally-regulated human subjects research, because they work with public data that does not (typically) contain personally-identifiable or other sensitive information. Nevertheless, we acknowledge that although we draw our dataset from a public source (publicly-available Reddit comments), there are still potential ethical concerns with using public data for research in ways that is unexpected by or potentially harmful to the people whose data is being used~\cite{fiesler}. Though the population of users that we study here, and the topics of the subreddits we study, are not particularly vulnerable or sensitive, we nevertheless take several precautions. In the paper, we do not name specific users or link to specific posts or comments, instead using anonymized user identifiers. When we make our labeled dataset publicly available, we will likewise exclude reddit username information. Finally, we did our best to remove posts and comments from our dataset (collected with ArcticShift) that were since deleted from the live Reddit site. We also verified that neither r/RealOrAI nor r/isthisAI have community guidelines in place that prohibit academic research from being conducted.

\item There are also ethical considerations with using LLMs to label data, since this may (for example) result in the unexpected downstream use of sensitive or participant data for model training. We selected the option to disallow Claude from using our data for training, and we used Anthropic's API to batch-send data, which (according to the company's documentation) is not used for training or improving models. Moreover, it is likely that public Reddit data has already been ingested by many companies for model training. 

\item In terms of adverse impacts, we recognize that AI companies or AI-generated content creators who are acting in bad faith (or simply want to improve the ``quality'' of their content by some metrics) could use our findings to make AI-generated content harder to detect. However, this is an ongoing arms race that is likely to be minimally impacted by our paper. Moreover, the data upon which we base our analyses are all already public on Reddit. Ultimately, we hope that this work will have positive impacts that outweigh its contributions to this arms race, in laying a foundation for future work that improves tools and education for consumers.

\end{enumerate}

\section{Author Positionality Statement}

All of the co-authors of this paper are based in the United States (one immigrated as a child from Western Europe), where our outlook is shaped by Western cultures and Western media systems.
Of the four co-authors, none 
of them are active community members (i.e., moderate or post or comment) on either r/RealOrAI or r/isthisAI, nor on any similar communities. The authors' combined areas of research experience cover computer security and privacy (including usable security), internet measurement, online communities, mis/disinformation, and perceptions of synthetic media.

%% file: sections/X_appendix.tex
\section{Additional Methodological Details}

\subsection{Prompt for Data Labeling}
When using Claude to classify comments, we used a single prompt to instruct the LLM to behave as a qualitative coder would, and tag each comment with as many categories as was appropriate. We provided a short description for each of the categories in our codebook, and also provided a category to represent when comments were not relevant to determining whether something was AI-generated or not (i.e., IRRELEVANT). We also provided specific instructions for Claude to follow, along with a defined JSON schema that included each label and a brief excerpt supporting each categorization from the comment.

Below, we provide the text of our final prompt.

\begin{tcolorbox}[
  colback=gray!5,
  colframe=gray!50,
  title=Classification Prompt,
  fonttitle=\bfseries,
  fontupper=\ttfamily\small,
  breakable,
  left=6pt,
  right=6pt
]
\textbf{\#\#\# TASK}\\
You are an expert at analyzing online comments to classify how people determine whether content is AI-generated or real.
Your task is to classify each input (representing one comment) into one or more of the following categories found outlined in the \#\#CATEGORY TAXONOMY section. When classifying each input, make sure to follow all the instructions outlined in the \#\#INSTRUCTIONS section. When available, examples provided in category sections are not exhaustive.

\vspace{6pt}
\noindent\rule{\linewidth}{0.4pt}
\vspace{4pt}

\textbf{\#\# CATEGORY TAXONOMY}\\[4pt]
\textbf{\#\#\# 1. MEDIA-BASED SIGNALS}\\
Signals drawn from examining the content itself.\\[4pt]
\textbf{\#\#\#\# A. Expectations of AI}
\begin{itemize}[leftmargin=*, nosep]
  \item \textbf{AI\_STYLE}: Comments explicitly identifies and discusses tells, hallmarks, characteristics, styles, that are indicative of AI
  \item \textbf{MODEL\_CAPABILITIES}: References beliefs about what AI can/can't do (e.g., "AI can't do physics," "Sora can only generate 15 seconds")
  \item \textbf{MODEL\_PROMPT\_SPECULATION}: Speculates on which model or prompt was used to generate the content, or hypothesizes that prompts were used to create the content
\end{itemize}

\textbf{\#\#\#\# B. Expectations of Reality}
\begin{itemize}[leftmargin=*, nosep]
  \item \textbf{PHYSICAL\_DETAILS}: Examines specific low-level physical elements in the content (e.g., lighting, shadows, finger count, object placement, or fine surface details or texture) to assess spatial or temporal consistency. Temporal signals include objects that change properties unnaturally over time (e.g., disappearance, object morphing).
  \item \textbf{CONTENT\_PLAUSIBILITY}: Signals drawn from high-level reasoning about whether the content makes sense as a real-world event -- without necessarily inspecting specific physical details. Includes questioning whether the scene is narratively plausible (e.g., "no one would film this"), whether the situation is contextually believable, or whether the content matches the commenter's broader lived experience or common sense.
  \item \textbf{MOTIVATION\_FOCUSED}: Questions \textit{why} content was made (e.g., political agenda, scam bait, engagement farming).
  \item \textbf{AI\_TRENDS}: Identifies the content as belonging to a broader online trend or genre that is commonly associated with AI-generated media -- e.g., recognizing it as part of a wave of AI-generated videos, a recurring meme format known to use AI, or a content category that has become a known AI production pattern (e.g., "these fake celebrity interview clips are all over the place lately").
  \item \textbf{INTUITION}: The comment attempts to take a position on whether content is AI-generated or human-made, but does not contain enough clear evidence, or uses vague gut feeling with no specific reasoning (e.g., "something feels off," "this is obviously AI").
\end{itemize}

\textbf{\#\#\# 2. METADATA / EXTERNAL SIGNALS}\\
Signals drawn from outside the content itself.
\begin{itemize}[leftmargin=*, nosep]
  \item \textbf{IDENTIFY\_SOURCE}: Identifies the original creator or channel or triangulates across multiple sources to establish provenance
  \item \textbf{CORROBORATING\_EVIDENCE}: Identifies and cites external information (news articles, other videos, links) to verify or debunk the content. DOES NOT use personal anecdotes as corroboration -- must be third party sources
  \item \textbf{LACK\_OF\_SOURCE}: Treats the absence of trustworthy, verifiable sources for the content as a signal that it is AI-generated -- e.g., no credible outlet has covered the event, no original upload can be traced, or the claim cannot be corroborated anywhere
  \item \textbf{MEDIA\_AGE}: Uses the age of the content as a signal -- e.g., content that predates widely accessible AI tools is more likely real; recently uploaded content may be more suspect
  \item \textbf{TOOL\_SIGNALS}: Comment explicitly discusses using external tools or methods to assess whether content is AI-generated, including, but not limited to:
    \begin{itemize}[nosep]
      \item \textit{Watermark detection}: References embedded signals such as SynthID or other provenance markers (e.g., watermarks, AI company logos)
      \item \textit{AI detection tools}: Uses dedicated software or services (e.g.\ GPTZero) designed to classify content as AI
      \item \textit{LLM queries}: Directly asks a large language model to assess or classify the content
      \item \textit{Reverse image/video search}: Traces content back to an original source or checks for prior appearances
      \item \textit{Copyright detection}: Flags content based on licensing or ownership metadata
    \end{itemize}
  \item \textbf{OTHER\_EXTERNAL}: Uses any other external signal not captured above to inform a judgment about whether content is AI-generated -- e.g., platform context, community reputation, or other circumstantial information
\end{itemize}

\textbf{\#\#\# 3. NON-CLASSIFICATION LABELS}
\begin{itemize}[leftmargin=*, nosep]
  \item \textbf{IRRELEVANT}: The comment does not take a position on whether the content is AI-generated or human-made (e.g., reacting to the content itself, asking unrelated questions, general conversation).
\end{itemize}

\vspace{6pt}
\noindent\rule{\linewidth}{0.4pt}
\vspace{4pt}

\textbf{\#\# INSTRUCTIONS}
\begin{enumerate}[nosep]
  \item Read the comment carefully.
  \item Assign one or more category labels from the taxonomy above.
  \item Quote the specific phrase(s) from the comment that support each label.
  \item Each label must have at least one supporting quote. A quote may be reused across labels if necessary.
  \item If the label is IRRELEVANT, return an empty evidence array.
  \item Return only valid JSON, as formatted in \#\#OUTPUT FORMAT. Do not include any explanation, preamble, or markdown formatting outside of the JSON object.
\end{enumerate}

\vspace{6pt}
\noindent\rule{\linewidth}{0.4pt}
\vspace{4pt}

\textbf{\#\# OUTPUT FORMAT}\\
Return your answer as a JSON object using this exact structure:\\[4pt]
\{\\
\hspace*{1em} "labels": ["LABEL\_1", ``LABEL\_2''],\\
\hspace*{1em} "evidence": [\\
\hspace*{2em}\{\\
\hspace*{3em} "label": "LABEL\_1",\\
\hspace*{3em} "quote": "exact quote from the comment"\\
\hspace*{2em}\},\\
\hspace*{2em}\{\\
\hspace*{3em} "label": "LABEL\_2",\\
\hspace*{3em} "quote": "exact quote from the comment"\\
\hspace*{2em}\}\\
\hspace*{1em}]\\
\}
\end{tcolorbox}

\subsection{Inter-Rater Reliability}

\input{tables/irr_combined_table}

In Table~\ref{tab:combinedirr}, we show the inter-rater reliability scores for several comparisons:

\subsubsection{Human IRR}
In Column 2, we show the inter-rater reliablity scores from two collaborators who categorized 100 comments. For any strategies with a $\kappa$ $\leq$ 0.7, the coders met to resolve disagreements.

\subsubsection{Prompt Development IRR}
In Column 3, we show the inter-rater reliability scores from human labeling compared to Claude labeling on the development set of 1,000 comments used while changing and updating the prompt. Scores are calculated based on the final prompt used.

\subsubsection{Holdout IRR}
In Column 4, we show the inter-rater reliability scores from human labeling against Claude labeling on the holdout set of 1,000 comments that were not used for prompt development, to ensure that we did not overfit the prompt when crafting the final version.

\begin{figure*}[tb]
    \centering
    \includegraphics[width=\textwidth]{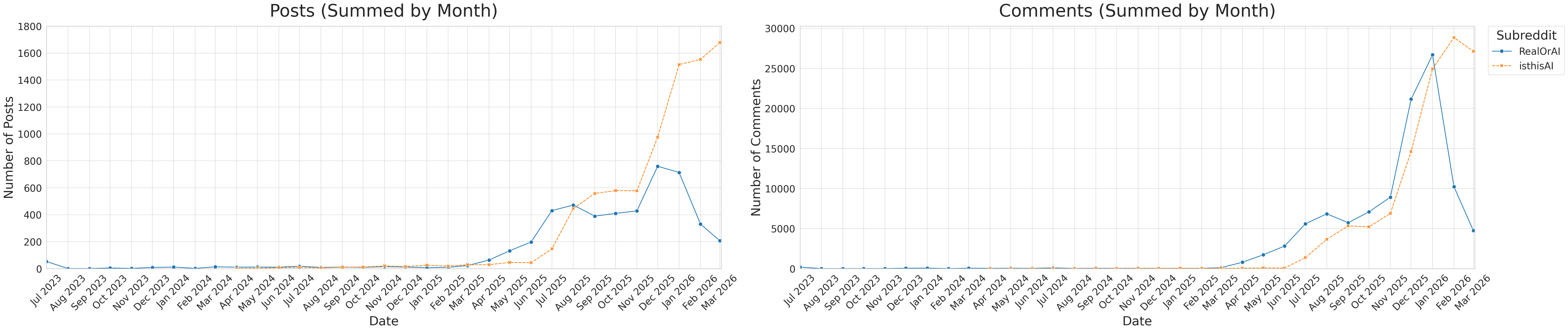}
    \caption{Posts and comments in our dataset from r/RealOrAI and r/isthisAI, from July 7, 2023 to March 31, 2026.}
    \label{fig:subreddit_overview}
\end{figure*}

\begin{figure}[tb]
    \centering
     \includegraphics[width=0.7\columnwidth]{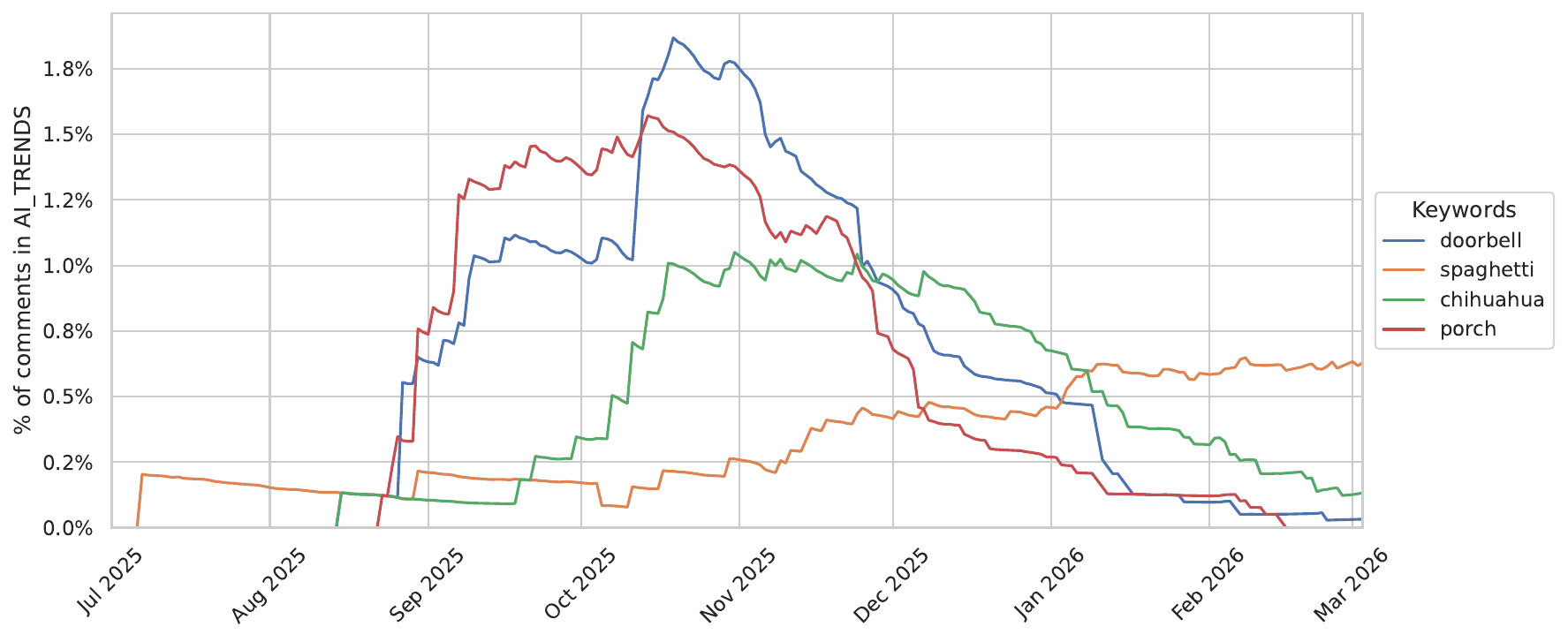}
     \caption{Centered average (over 90 days) of comments in the AI\_TRENDS sub-code that uses specific keywords. Words such as ``doorbell'' appear to have distinct temporal spikes (around the Halloween period), while others like ``spaghetti'' are used relatively consistently.}
     \label{fig:ai_trends_keywords}
 \end{figure}

\section{Additional Findings}

\subsection{Dataset Over Time}

Figure~\ref{fig:subreddit_overview} shows the number of posts and comments in our dataset over time, from both subreddits.

\subsection{Additional Word Frequency Data}

Table~\ref{tab:tfidf_appendix} shows the 10 most common unique words for six sub-codes, based on tf-idf.

\input{tables/appendix_tfidf}

\subsection{AI Trends Over Time}
Figure \ref{fig:ai_trends_keywords} shows how top keywords in comments labeled AI\_TRENDS vary over time as trends come and go: \textit{doorbell, spaghetti, chihuahua,} and \textit{porch}.

\input{sections/X_1_political}

%% file: tables/irr_combined_table.tex
\begin{table*}[tb]
\centering
\small
\begin{tabular}{@{}lccc@{}}
\toprule
\textbf{Category or Sub-code}       & \textbf{Human IRR} & \textbf{Prompt Development IRR} & \textbf{Holdout IRR} \\ \midrule
\textbf{EXTERNAL\_SIGNALS}                   &            &            &            \\ \midrule
TOOL\_SIGNALS              & $\kappa$ = 1.00 & $\kappa$ = 0.71 & $\kappa$ = 0.93 \\
TRIANGULATING\_INFO        & $\kappa$ = 1.00 & $\kappa$ = 0.89 & $\kappa$ = 0.90 \\
MEDIA\_AGE                 & $\kappa$ = 0.76 & $\kappa$ = 0.96 & $\kappa$ = 0.96 \\
OTHER\_EXTERNAL            & $\kappa$ = 1.00 & $\kappa$ = 0.74 & $\kappa$ = 1.00 \\ \midrule
\textbf{CONTENT\_FOCUSED}                    &            &            &            \\ \midrule
AI\_STYLE                  & $\kappa$ = 1.00 & $\kappa$ = 0.89 & $\kappa$ = 0.84 \\
MODEL\_CAPABILITIES        & $\kappa$ = 0.92 & $\kappa$ = 0.91 & $\kappa$ = 0.87 \\
MODEL\_PROMPT\_SPECULATION & $\kappa$ = 1.00 & $\kappa$ = 0.97 & $\kappa$ = 0.85 \\
CONTENT\_PLAUSIBILITY      & $\kappa$ = 0.97 & $\kappa$ = 0.93 & $\kappa$ = 0.95 \\
PHYSICAL\_DETAILS          & $\kappa$ = 1.00 & $\kappa$ = 0.97 & $\kappa$ = 0.99 \\
INTUITION                  & $\kappa$ = 0.82 & $\kappa$ = 0.86 & $\kappa$ = 0.85 \\
AI\_TRENDS                 & $\kappa$ = 1.00 & $\kappa$ = 0.93 & $\kappa$ = 0.93 \\
MOTIVATION\_FOCUSED        & $\kappa$ = 1.00 & $\kappa$ = 0.71 & $\kappa$ = 0.86 \\ \bottomrule
\end{tabular}
\caption{Cohen's kappa inter-rater reliability scores for each strategy.}
\label{tab:combinedirr}
\end{table*}

%% file: tables/appendix_tfidf.tex
\begin{table*}[tb]
\caption{Top 10 unique keywords in specific subcodes (resulting from tf-idf). Words are ordered in decreasing tf-idf value.}
\small \centering
\label{tab:tfidf_appendix}
\begin{tabular}{@{}llllll@{}}
\toprule
\begin{tabular}[c]{@{}l@{}}CONTENT\\ PLAUSIBILITY\end{tabular} & INTUITION            & \begin{tabular}[c]{@{}l@{}}MODEL\\ CAPABILITIES\end{tabular} & \begin{tabular}[c]{@{}l@{}}OTHER\\ EXTERNAL\end{tabular} & \begin{tabular}[c]{@{}l@{}}PHYSICAL\\ DETAILS\end{tabular} & \begin{tabular}[c]{@{}l@{}}TRIANGULATE\\ INFO\end{tabular} \\ \midrule
torch                                                          & fakest               & accurately                                                   & vgen                                                     & morphs                                                     & bbc                                                        \\
\rowcolor[HTML]{EFEFEF} 
rats                                                           & risthisaicirclejerk & nespresso                                                    & january                                                  & torch                                                      & national                                                   \\
react                                                          & painfully            & permanence                                                   & crypto                                                   & handles                                                    & maps                                                       \\
\rowcolor[HTML]{EFEFEF} 
resin                                                          & christ               & readable                                                     & youtubes                                                 & pupils                                                     & san                                                        \\
intubated                                                      & faker                & diffusion                                                    & raivideo                                                 & stripes                                                    & passenger                                                  \\
\rowcolor[HTML]{EFEFEF} 
flour                                                          & 100000               & brushes                                                      & recommended                                              & flour                                                      & wildlife                                                   \\
pour                                                           & cybyapi              & react                                                        & reputation                                               & tiles                                                      & january                                                    \\
\rowcolor[HTML]{EFEFEF} 
nasal                                                          & humanity             & maintain                                                     & rife                                                     & elbow                                                      & admitted                                                   \\
cannula                                                        & soulless             & generates                                                    & neveah                                                   & gap                                                        & 2016                                                       \\
\rowcolor[HTML]{EFEFEF} 
sink                                                           & 1000000              & manage                                                       & fades                                                    & index                                                      & broke                                                      \\ \bottomrule
\end{tabular}
\end{table*}

%% file: sections/X_1_political.tex
\begin{figure}[tb]
    \centering
    \vspace{2in}
    \includegraphics[width=0.7\columnwidth]{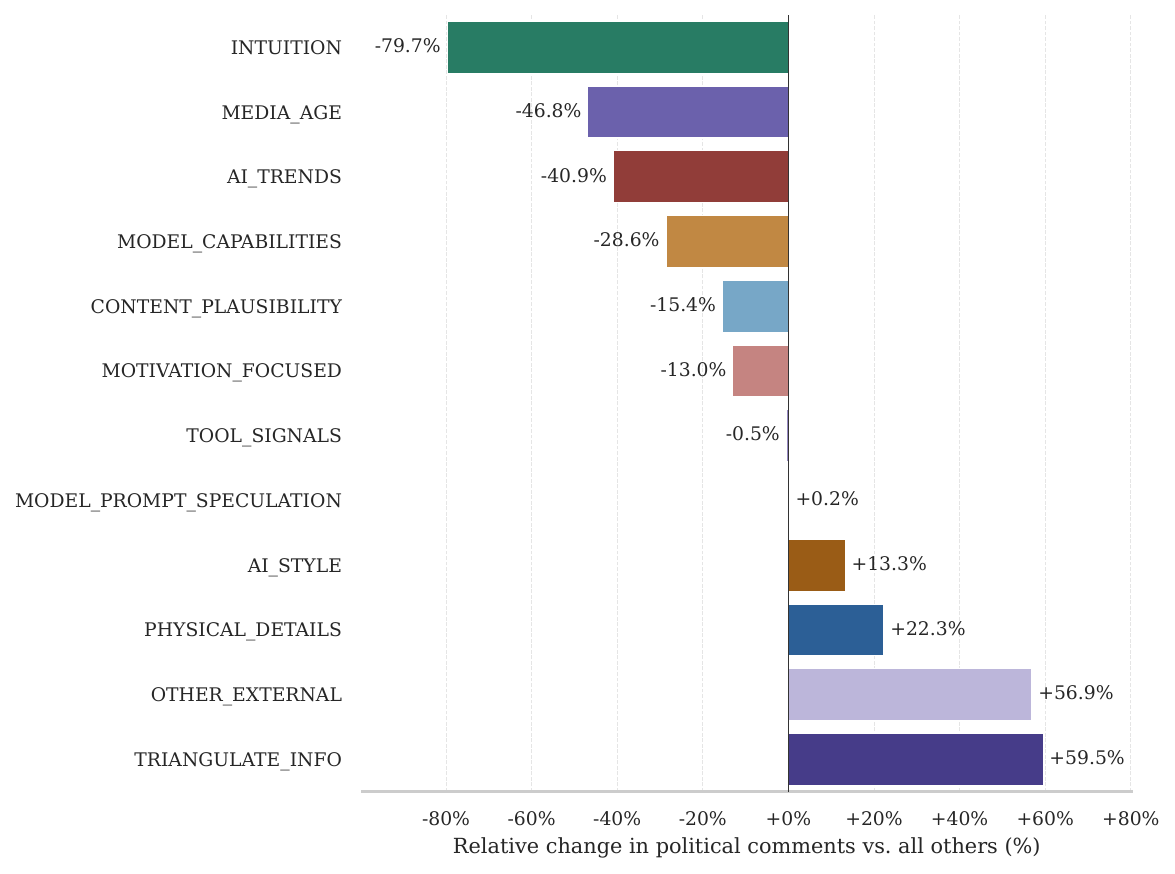}
    \caption{The relative difference in strategies in comments discussing political-related posts, versus all other posts from December 2025 to March 2026. Political posts are 59.5\% more likely to use TRIANGULATE\_DETAILS, 22.3\% more likely to use PHYSICAL\_DETAILS.}
    \label{fig:politics_barchart}
\end{figure}

\subsection{Strategies for Political vs. Non-Political Content}
On r/isthisAI, moderators require users to label posts about political content with the ``Politics'' flair~\cite{politics-moderation}. 
Though comments on these posts represent a small portion of our dataset (772 comments),
the labeled data enables us to compare the strategies that people describe when commenting on political content to the strategies they use for other content (i.e., posts without the ``Politics'' flair).
Posts labeled with the ``Politics'' flair begin only in January 2026, so we limit our analysis of political vs. non-political content to December 2025 -- March 2026 (end of our dataset). Focusing on this time period, we compare the 772 comments on political posts with 121,857 comments on other posts.

To examine whether the strategies that people describe when analyzing political content differ, we compare the relative prevalence of the strategies (from our codebook) in each slice of the data. 
Figure \ref{fig:politics_barchart} shows these relative differences for comments on posts with the ``Politics'' flair and posts without. While $\chi^2$ tests show that the \textit{overall} distributions between political and all other comments are not statistically significantly different, our post-hoc investigation of \textit{specific} strategies reveals some differences (Table~\ref{tab:posthoc_politics}).

\input{tables/post_hoc_politics}

We find that relative to non-political posts, people appear to use PHYSICAL\_DETAILS and TRIANGULATING\_DATA (statistically significantly) more often than when evaluating all other types of content in r/isthisAI. Conversely, we see that people appear to use INTUITION and CONTENT\_PLAUSIBILTIY (statistically significantly) less.
Referring back to Figure~\ref{fig:politics_barchart}, for example, 
PHYSICAL\_DETAILS is used 22.3\% more often for political posts.
Though our methodology cannot explain the root causes of these differences, our findings suggest that users gravitate to different strategies for different types of content.

%% file: tables/post_hoc_politics.tex
\begin{table*}[tb]
\centering
\footnotesize
\caption{Chi-squared post-hoc results, comparing strategies used in politics-related comments and all other comments in r/isthisAI from December 2025 to March 2026. Bolded rows represent statistic significance, post-Bonferroni corrections.}
\label{tab:posthoc_politics}
\begin{tabular}{llllrl}
\hline
\textbf{Sub-code}          & \textbf{Politics} & \textbf{All Others} & \textbf{$\chi^2$} & \multicolumn{1}{l}{\textbf{Delta}} & \textit{\textbf{p-val}} \\ \hline
\textbf{PHYSICAL\_DETAILS} & \textbf{576}      & \textbf{62,969}     & \textbf{35.965}   & \textbf{22.3\%}                    & \textbf{0.000}          \\
\rowcolor[HTML]{EFEFEF} 
CONTENT\_PLAUSIBILITY      & 254               & 40,150              & 9.001             & -15.4\%                            & 0.032                   \\
AI\_STYLE                  & 148               & 17,472              & 2.385             & 13.3\%                             & 1.000                   \\
\rowcolor[HTML]{EFEFEF} 
MODEL\_CAPABILITIES        & 42                & 7,865               & 4.715             & -28.6\%                            & 0.359                   \\
MODEL\_PROMPT\_SPECULATION & 28                & 3,738               & 0.000             & 0.2\%                              & 1.000                   \\
\rowcolor[HTML]{EFEFEF} 
\textbf{TRIANGULATE\_INFO} & \textbf{148}      & \textbf{12,407}     & \textbf{34.311}   & \textbf{59.5\%}                    & \textbf{0.000}          \\
MEDIA\_AGE                 & 17                & 4,276               & 6.691             & -46.8\%                            & 0.116                   \\
\rowcolor[HTML]{EFEFEF} 
TOOL\_SIGNALS              & 18                & 2,419               & 0.000             & -0.5\%                             & 1.000                   \\
OTHER\_EXTERNAL            & 9                 & 767                 & 1.318             & 56.9\%                             & 1.000                   \\
\rowcolor[HTML]{EFEFEF} 
AI\_TRENDS                 & 17                & 3,847               & 4.493             & -40.9\%                            & 0.409                   \\
MOTIVATION\_FOCUSED        & 29                & 4,455               & 0.445             & -13.0\%                            & 1.000                   \\
\rowcolor[HTML]{EFEFEF} 
\textbf{INTUITION}         & \textbf{22}       & \textbf{14,520}     & \textbf{74.281}   & \textbf{-79.7\%}                   & \textbf{0.000}          \\ \hline
\end{tabular}
\end{table*}

